\documentclass[english,aps,prl,showpacs,superscriptaddress,floats,amsmath,amssymb,floatfix,twocolumn,nobalancelastpage]{revtex4-1}
\usepackage[T1]{fontenc}
\usepackage[latin9]{inputenc}
\usepackage{color}
\usepackage{textcomp}
\usepackage{amsmath}
\usepackage{amssymb}
\usepackage{graphicx}
\usepackage{esint}
\usepackage{wasysym}
\usepackage{lipsum}
\usepackage{hyperref}
\usepackage{babel}

\def\XXint#1#2#3{{\setbox0=\hbox{$#1{#2#3}{\int}$}
     \vcenter{\hbox{$#2#3$}}\kern-.5\wd0}}

\begin{document}

\title{Heat transport in the anisotropic Kitaev spin liquid} 

\author{Angelo Pidatella}

\email{angelo.pidatella@tu-dresden.de}

\selectlanguage{english}%

\affiliation{Institute for Theoretical Physics, Technical University Dresden,
D-01062 Dresden, Germany}

\author{Alexandros Metavitsiadis}

\email{a.metavitsiadis@tu-bs.de}

\selectlanguage{english}%

\affiliation{Institute for Theoretical Physics, Technical University
Braunschweig, D-38106 Braunschweig, Germany}

\author{Wolfram Brenig}

\email{w.brenig@tu-bs.de}

\selectlanguage{english}%

\affiliation{Institute for Theoretical Physics, Technical University
Braunschweig, D-38106 Braunschweig, Germany}

\date{\today}
\begin{abstract}
We present a study of longitudinal thermal transport in the Kitaev spin model on
the honeycomb lattice, focusing on the role of anisotropic exchange to cover
both, gapless and gapped phases.  Employing a complementary combination of exact
diagonalization on small systems and an average gauge configuration approach for
up to $\sim O(10^4)$ spinful sites, we report results for the dynamical energy
current auto-correlation function as well as the dc thermal conductivity over a
wide range of temperatures and exchange anisotropies. Despite a pseudogap in the
current correlation spectra, induced by emergent thermal gauge disorder on any
finite system, we find that in the thermodynamic limit, gapless and gapful
phases both feature normal dissipative transport, with a temperature dependence
crossing over from power law to exponentially activated behavior upon gap
opening.
\end{abstract}

\maketitle

\section{Introduction}

\noindent

Quantum spin liquids (QSL) represent a rare form of magnetic matter, with
fluctuations strong enough to suppress the formation of local order parameters
even at zero temperature. QSLs typically result from frustration of magnetic
exchange and tend to exhibit many peculiar properties, such as massive
entanglement, quantum orders and fractionalized excitations \cite{Balents2010,
Balents2016}. Among the many models proposed for putative QSLs, Kitaev's compass
exchange Hamiltonian on the honeycomb lattice stands out as one of the few, in
which a $\mathbb{Z}_{2}$ QSL can exactly be shown to exist \cite{Kitaev2006}.
In fact, pairs of spins in this model fractionalize in terms of mobile Majorana
fermions coupled to static $\mathbb{Z}_{2}$ gauge fields
\cite{Kitaev2006,Feng2007, Chen2008, Nussinov2009, Mandal2012}. Solid state
realizations of Kitaev's model are based on Mott-insulators with strong
spin-orbit coupling (SOC) \cite{Khaliullin2005, Jackeli2009,Chaloupka2010,
Nussinov2015}, although residual non-Kitaev exchange interactions remain an
issue at low temperatures and low energies, where most of the present systems
still tend to order magnetically \cite{Trebst2017}.

Mobile Majorana matter has been suggested to leave several fingerprints in
spectroscopic measurements, like inelastic neutron
\cite{Banerjee2016,Banerjee2016a, Banerjee2018}, Raman scattering
\cite{Knolle2014}, as well as in local resonance probes \cite{Baek2017,
Zheng2017}. Apart from spectroscopy, thermal transport is yet another powerful
tool, able to discriminate microscopic models of elementary excitations and
their scattering mechanisms in quantum magnets - see Ref. \cite{Hess2018a} for a
review. Among the candidates for Kitaev QSLs, $\alpha$-$\mathrm{RuCl_{3}}$
\cite{Plumb2014} has been under intense scrutiny regarding measurements of the
longitudinal thermal conductivity $\kappa$ \cite{Hirobe2017, Leahy2017,
Hentrich2018, Yu2018}. At first, heat transport by itinerant spin excitations
had been inferred from an anomaly in the in-plane longitudinal heat conductivity
$\kappa_{xx}$ at around 100~K \cite{Hirobe2017}, and from a magnetic
field-induced low-temperature enhancement of $\kappa_{xx}$ for fields $B\gtrsim
8$~T parallel to the material's honeycomb planes \cite{Leahy2017}. These
interpretations however, have recently been ruled out by results for the
out-of-plane heat conductivity $\kappa_{zz}$ where both types of anomalies are
present as well \cite{Hentrich2018}. The emergent picture for rationalizing the
{\it longitudinal} heat transport in $\alpha$-$\mathrm{RuCl_{3}}$ material is
thus, that its most salient features can be explained by phononic heat transport
which is likely to be renormalized by scattering from Majorana matter
\cite{Hentrich2018,Yu2018}. However, it remains to be settled whether also a
sub-leading direct magnetic contribution to the heat transport is present,
masked by the phononic transport. Very recently also {\it transverse} thermal
transport in magnetic fields, i.e. thermal Hall conductivity $\kappa_{xy}$, and
its potential quantization has been observed in $\alpha$-$\mathrm{RuCl_{3}}$
\cite{Kasahara2018}. This may be a rather strong evidence for Kitaev physics and
chiral Majorana edge modes in this material.

Theoretically, thermal transport studies in pure Kitaev QSLs have been performed
using quantum Monte Carlo in 2D \cite{Motome2017} or applying ED in 1D and 2D
\cite{Steinigeweg2016, Metavitsiadis2016, Metavitsiadis2017}. Thermal transport
in Kitaev-Heisenberg ladders has been considered recently, employing ED and
quantum typicality \cite{Metavitsiadis2018a}. Magnetically {\it ordered} phases
of a Kitaev-Heisenberg model have also been investigated for transport using
spin wave calculations \cite{Stamokostas2017}. Remarkably, in the 2D QSL
case, all of these studies have been confined to the point of
C$_3$-symmetrically sized, isotropic compass exchange, i.e. the {\it gapless}
phase of the Kitaev model. The main purpose of this work is to step beyond
this limitation and investigate the longitudinal thermal conductivity, covering
also anisotropic Kitaev exchange, ranging from the gapless to the gapped
case.  Exploring both phases, we find that regardless of the exchange coupling 
regime, the system supports a finite dc transport coefficient at all {\it non-zero}
temperatures investigated, however with the emergence of thermal activation
behavior in the gapped phase.

The outline of this paper is as follows: in Sec.~\ref{secModel} we briefly
recall the properties of the Kitaev model, along with basic ingredients of
thermal transport calculations, as used in this work. In Sec.~\ref{secResults}
we detail our results for the thermal conductivity versus the anisotropy of the
exchange coupling, as obtained using the average gauge configurations (AGC). The
method is explained in Sec.~\ref{secAGC}, which is then applied to study the dynamical heat current
auto-correlation function in Sec.~\ref{sub-secHeatCorr}, and the dc heat
conductivity in Sec.~\ref{sub-secDCkappa}.  We conclude in
Sec.~\ref{secConcl}. Appendix \ref{appKM} relates our findings to
transport-phenomenology.
 
\section{Model}\label{secModel}

\begin{figure}[tb]
\centering{}\includegraphics[width=0.95\columnwidth]{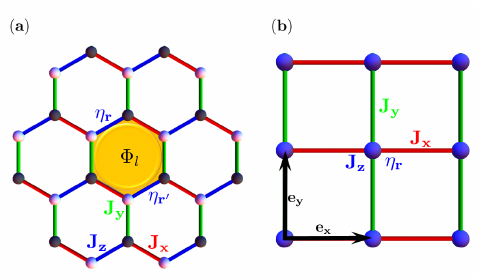}
\caption{ (a) Kitaev model on honeycomb lattice
(HL). Red, green, blue colors label $J_{x,y,z}$ exchange interactions between
black and white triangular sublattice sites. Yellow $\Phi_{l}=\eta_{\bf
r}\eta_{{\bf r}'}$ refers to the flux, with gauge fields $\eta_{\bf r},\eta_{{\bf
r}'}$ attached to blue edge of the hexagon.  (b) Kitaev model, shrunk onto
square lattice (SL) of Eq.~(\ref{eq:a2}). $J_{x,y}$ (red, green edges) mediate
fermion hopping/pairing, along lattice vectors $\mathbf{e_{x,y}}$. Exchange
interactions $J_{z}$ and gauge fields $\eta_{\bf r}$ (blue vertices) set on-site
energy.}
\label{fig:latticeSQ}
\end{figure}

We study the longitudinal dynamical thermal conductivity
of the Kitaev model in the presence of \emph{anisotropy}, introduced by tuning the
relative exchange couplings and considering both, gapless and gapped regimes. The
Hamiltonian is
\begin{equation}
\mathcal{H}=\sum_{\langle l,m\rangle,\lambda(\langle l,m\rangle)}J_{\lambda}
\sigma_{l}^{\lambda}
\sigma_{m}^{\lambda}\,,\label{eq:a1}
\end{equation}
where $\langle l,m \rangle$ indicates nearest-neighbors on a 2D honeycomb
lattice (HL) of 2$N$ sites, with $J_{\lambda}$ and $\sigma^{\lambda}$ the
exchange couplings and the Pauli matrices, respectively, for coordinates
$\lambda=x,y,z$, as shown in Fig.~\ref{fig:latticeSQ}(a). Model (\ref{eq:a1})
can be mapped onto a sum of $2^N$ Hamiltonians of non-interacting itinerant
Majorana fermions, each of which is classified by a fixed, distinct distribution
of values of $N$ static $\mathbb{Z}_{2}$ gauge fields $\eta_{\bf r}=\pm 1$,
residing on, e.g., the blue bonds in Fig. \ref{fig:latticeSQ}(a). The Majorana
fermions of each gauge sector $\{\eta_{\bf r}\}$ can be rewritten in terms of
$N$ spinless non-interacting fermions $d_{\mathbf{r}}^{(\dagger)}$ with
occupation numbers 0,1. This implies a total of $2^{2N}$ states of gauges and
spinless fermions, consistent with the dimension of the original spin Hilbert
space. Each of the $2^N$ fermionic sub-Hamiltonians is a QSL with spin correlations
not exceeding nearest neighbor distance. Several routes have been documented
to achieve the aforementioned mapping and we refer to Refs.
\cite{Kitaev2006, Feng2007, Chen2008, Nussinov2009, Mandal2012}
for details. In this $\eta_{\bf r},
d_{\mathbf{r}}^{(\dagger)}$-language (\ref{eq:a1}) reads
\begin{eqnarray}
\mathcal{H} && = \sum_{\mathbf{r}}h(\mathbf{r}) =
\sum_{\mathbf{r}}\left[ J_{x}(d_{\mathbf{r}}^{\dagger}+
d_{\mathbf{r}}^{\phantom{\dagger}})(d_{\mathbf{r}+\mathbf{e}_{x}}^{\dagger}
-d_{\mathbf{r}+\mathbf{e}_{x}}^{\phantom{\dagger}})+\right.
\nonumber \\
&&\hphantom{a}
\left. J_{y}(d_{\mathbf{r}}^{\dagger}+d_{\mathbf{r}}^{\phantom{\dagger}})
(d_{\mathbf{r}+\mathbf{e}_{y}}^{\dagger}-d_{\mathbf{r}+
\mathbf{e}_{y}}^{\phantom{\dagger}})+J_{z}\eta_{\mathbf{r}}(
2d_{\mathbf{r}}^{\dagger}d_{\mathbf{r}}^{\phantom{\dagger}}-1)\right]\, .
\hphantom{a}
\label{eq:a2}
\end{eqnarray}
Here, for convenience, the $J_z$-bonds, which feature only an $\eta_{\bf
r}$-dependent on-bond potential for the fermions are shrunk onto just a point,
transforming the HL into a square lattice (SL), with primitive vectors ${\bf
e}_{x,y}$ as depicted in Fig.~\ref{fig:latticeSQ}(b). We will refer the SL
geometry throughout this paper. To quantify the {\it anisotropy} of the Kitaev
exchange, we employ a parameter $\alpha$, such that $J_{x}=J_{y}=\alpha$, and
$J_{z}=3-2\alpha$. Tuning $\alpha$ allows to keep the fermionic energy scale
$(J_x{+}J_y{+}J_z) {/} 3 {=} 1 {\equiv} J$ constant, while accessing both, the
gapless and the gapped regime of the zero temperature spectrum of (\ref{eq:a2}),
for $1>\alpha>0.75$ and $0.75>\alpha>0$, respectively \cite{Kitaev2006}.

Our discussion of the thermal conductivity is based on calculations within
linear response theory, assuming the long wave length limit $\mathbf{q}
{\rightarrow} 0$. For this, thermal transport coefficients are computed from the
energy current auto-correlation functions $\langle{\cal J}(t){\cal J}\rangle$,
where ${\cal J}$ is the energy current and $\langle\dots \rangle$ refers to the
thermal trace \cite{FHM2007}. The real part of thermal conductivity, i.e. $\kappa^{
\prime } ( \omega )$, follows from the Fourier transform
\begin{eqnarray}  
C(\omega)  &=&  \frac{1}{N}\int dt e^{i\omega t}
\langle{\cal J}(t){\cal J}\rangle , \label{eq:a3} \\
\kappa^{\prime}(\omega)  &=&  \frac{\beta}{2\omega} \left(
1-e^{-\beta\omega}\right)\, C(\omega)~.  \label{eq:a4} 
\end{eqnarray}
Generically, $\kappa^{\prime}(\omega) = 2\pi D \delta(\omega)+ \kappa^{reg} (
\omega )$. The {\it Drude} weight (DW) $D$ refers to a ballistic channel, and if
finite marks a perfect conductor. $\kappa^{reg} ( \omega )$ encodes the
dissipative parts of the heat flow. If $\kappa^{dc} = \kappa^{reg}( \omega
\rightarrow 0)$ is finite, the system is a normal heat conductor with a finite
DC conductivity. Interacting quantum systems with $D\neq 0$ are rare, see e.g. Ref.
\cite{FHM2007}, and $D=0$ for $N\rightarrow\infty$ has also been argued for 
the Kitaev model \cite{Metavitsiadis2016,Metavitsiadis2017}. In turn, a prime
goal of this work will be the analysis of $\kappa^{reg} ( \omega )$ and its
limit for $\omega\rightarrow 0$ in order to determine the DC conductivity.

To derive the heat current operator in (\ref{eq:a3}) one has to realize that
most of the gauge sectors in (\ref{eq:a2}) are not translationally invariant,
due to the real space distribution of $\eta_{\bf r}$. In turn, we employ the
polarization operator $\mathbf{P}$ \cite{Metavitsiadis2017}, to obtain the
current operator in real space
\begin{equation}
 \mathbf{P}=\sum_{\mathbf{r}}\mathbf{r}h(\mathbf{r}),\,\,\,{\cal
 J}=\frac{\partial
 \mathbf{P}}{\partial t}=
 i\left[{\cal H},\mathbf{P}\right]\, .
 \label{eq:a5}
\end{equation}
It is worth noting, that in a translationally invariant system ${\cal J}$ from
(\ref{eq:a5}) is exactly identical to that obtained from the continuity equation
in the limit of $\mathbf{q}{\rightarrow} 0$. On the SL we get
\begin{equation}
{\cal J}_{\mu} =2iJ_\mu \sum_{\mathbf{r}}
[ J_{z} \eta_{\mathbf{r}}\chi_{\mathbf{r}}\chi_{\mathbf{r}-\mathbf{e}_{\mu}}
+ \tau_\mu J_{\bar{\mu}}\chi_{\mathbf{r}}\chi_{\mathbf{r}+\mathbf{e}_{x}-
\mathbf{e}_{y}}]~, 
\label{eq:a6}
\end{equation}
with $\chi_{ \mathbf{r}}=(d_{ \mathbf{r}}^{ \dagger}{+}d_{ \mathbf{r}})$, $\bar
\mu = y(x)$ and $\tau_\mu{=}+(-)$ for $\mu=x(y)$.

For the remainder of this work we resort to numerical methods to treat
(\ref{eq:a3}). For this we note, that not only the Hamiltonian, but also the
current operator does not mediate transitions between gauge sectors. In turn,
any thermal trace can be decomposed into a classical trace over gauges and a
trace over the fermions, separately. For the latter, and for each fixed gauge
sector $\{\eta\}=(\eta_{1}, \dots, \eta_{N})$, we define a $2N$ component spinor
$\mathbf{D}^{\dagger} = ( d^{\dagger}_{1}, \dots, d^{\dagger}_{N}, d_{1}, \dots,
d_{N})$ of the fermions, with indices $1\dots N$ referring to the sites on the
SL. With this the Hamiltonian and the current for each gauge sector read
$\mathcal{H} = \mathbf{D^{\dagger} h (\eta) D}$, and ${\cal J}_{\mu} =
\mathbf{D^{\dagger} j_{\mu}(\eta) D}$. Using a Bogoliubov transformation $U$
onto quasiparticle fermions $\mathbf{A}^{\dagger} = (a^{\dagger}_{1}, \dots,
a^{\dagger}_{N}, a_{1}, \dots, a_{N})$ via $\mathbf{A = U^{\dagger} D}$, the
Hamiltonian is diagonalized as
\begin{equation}
 \mathcal{H}=\sum_{i}\varepsilon_{i}(\{\eta\})(a^{\dagger}_{i}a_{i}-\frac{1}{2}),\ 
 \label{eq:a7}
\end{equation}
with quasiparticle energies $\varepsilon_{i}(\{\eta\})$. The contribution of
each gauge sector to (\ref{eq:a3}) can then be obtained straightforwardly by a
Wick-decomposition of $\langle{\cal J}(t){\cal J}\rangle$ in the quasiparticle
basis \cite{Metavitsiadis2016}.

\section{Results}\label{secResults}

In this section we will present our results for the dynamical energy current
auto-correlation function $C(\omega,T)$ and the dc-limit of the heat
conductivity $\kappa^{dc}(T)$ versus anisotropy at finite temperatures.

\begin{figure}[tb]
\centering{}\includegraphics[width=0.95\columnwidth]{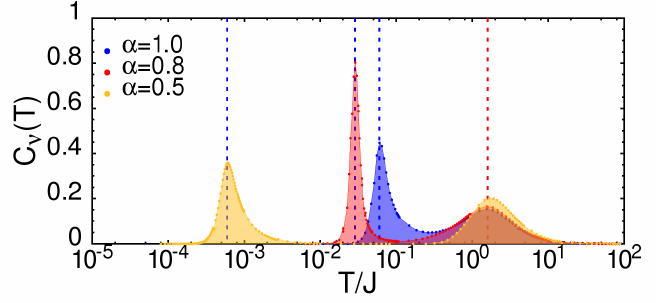}
\caption{ Specific heat $C_{\nu}(T)$ for $\alpha=1.0, 0.8, 0.5$ from ED on
$N{=}6{\times}6$, i.e. 72 spinful sites versus $T$. Vertical lines at $T^\star_\alpha$
(blue dashed) and $T_M$ (red dashed) indicate temperature scales for flux and
fermion entropy release.}
\label{fig:SHalpha1}
\end{figure}

\subsection{Average gauge configuration method}\label{secAGC}

Tracing over the gauge sectors in (\ref{eq:a3}) can be done in various ways and
to various degrees of precision. Among the methods are Markovian sampling of the
gauges by QMC \cite{Motome2015b, Motome2017}, exact summation over all gauges,
equivalent to ED of the spin model \cite{Metavitsiadis2016, Metavitsiadis2017},
and summing only over a dominant subset of gauge configurations, i.e. the
average gauge configuration (AGC) method \cite{Metavitsiadis2016,
Metavitsiadis2017}, which will be used here. The following sketches the  main ideas
of the AGC method, while for more details we refer the reader to Refs.~\cite{Metavitsiadis2016, Metavitsiadis2017}.  The notion of an average
gauge configuration is trivial in the low- and the high-temperature limit. In
the former, the gauges assume a flux-free configuration \cite{Kitaev2006}, and
in the latter a typical gauge configuration is fully random. These limiting
cases are separated by a temperature scale $T^\star_\alpha$ at which
flux-proliferation occurs upon increasing temperature. Here $\alpha$ refers to
the anisotropy parameter. $T^\star_\alpha$ scales with the energy to excite a
single flux, i.e. the flux gap $\Delta_\Phi/ J \ll 1$ \cite{Kitaev2006}. Because
of the latter inequality, $T^\star_\alpha$ and the crossover regime can be read
off from the specific heat, i.e. from the entropy release of the gauge fields
\cite{Motome2015b}, since this is well separated from the fermionic energy
scales $O(J)$. Our strategy will be to first determine $T^\star_\alpha$ and then
to confine all evaluations of $C(\omega)$ to temperatures above this scale, such
that averaging over only completely {\it random} gauge states is sufficiently
justified for the gauge trace. Consequently, the remaining effective fermionic
model is a binary disorder problem. It turns out that the temperature range
accessible by this approach is large enough for our purposes. \\

To approximate $T^\star_\alpha$ we now calculate the specific heat $C_{\nu}(T) =
T \partial S / \partial T$ \cite{noteSpec}. This is done {\it exactly} on finite
systems, tracing over both, fermions and gauges. The system sizes we can reach
for this, namely $N {\leq} 6 {\times} 6$, i.e. 72 spins, are larger than those
for exact thermodynamics using the many body {\it spin basis}, however, still
much smaller than those which will be used within the AGC
subsequently. Fig.~\ref{fig:SHalpha1} shows the temperature dependence of the
specific heat for three different values of anisotropy. As expected, two peaks
develop at two different temperature scales, corresponding to the release of
entropy from the gauge degrees of freedom and the mobile Majorana fermions, at
$T^{ \star }_{ \alpha }$ and $T_{M}/J{\sim}1$, respectively. Obviously, the
global fermionic energy scale $J$ is only weakly affected by the gauge disorder,
and therefore the high-$T$ peak remains rather insensitive to $\alpha$. However
the low-$T$ peak at $T^{ \star }_{ \alpha }$ is strongly shifted to smaller
temperatures, by roughly two orders of magnitude, upon decreasing $\alpha$ from
1 to 0.5. This finding is consistent with results from QMC
\cite{Motome2015b}. While in principle $T^{ \star }_{ \alpha }$ sets an
individual temperature scale for each $\alpha$, above which averaging over fully
random gauge configurations provides a sufficient approximation for the gauge
trace, we will use the {\it maximum} of these, i.e. $T^{ \star,\mathrm{max} }_{
\alpha }/J\sim O(.1)$ as the lowest temperature to apply the AGC for all
subsequent calculations, independent of $\alpha$.

\subsection{Dynamical current correlations}
\label{sub-secHeatCorr}

\begin{figure}[tb]
\centering{}\includegraphics[width=0.95\columnwidth]{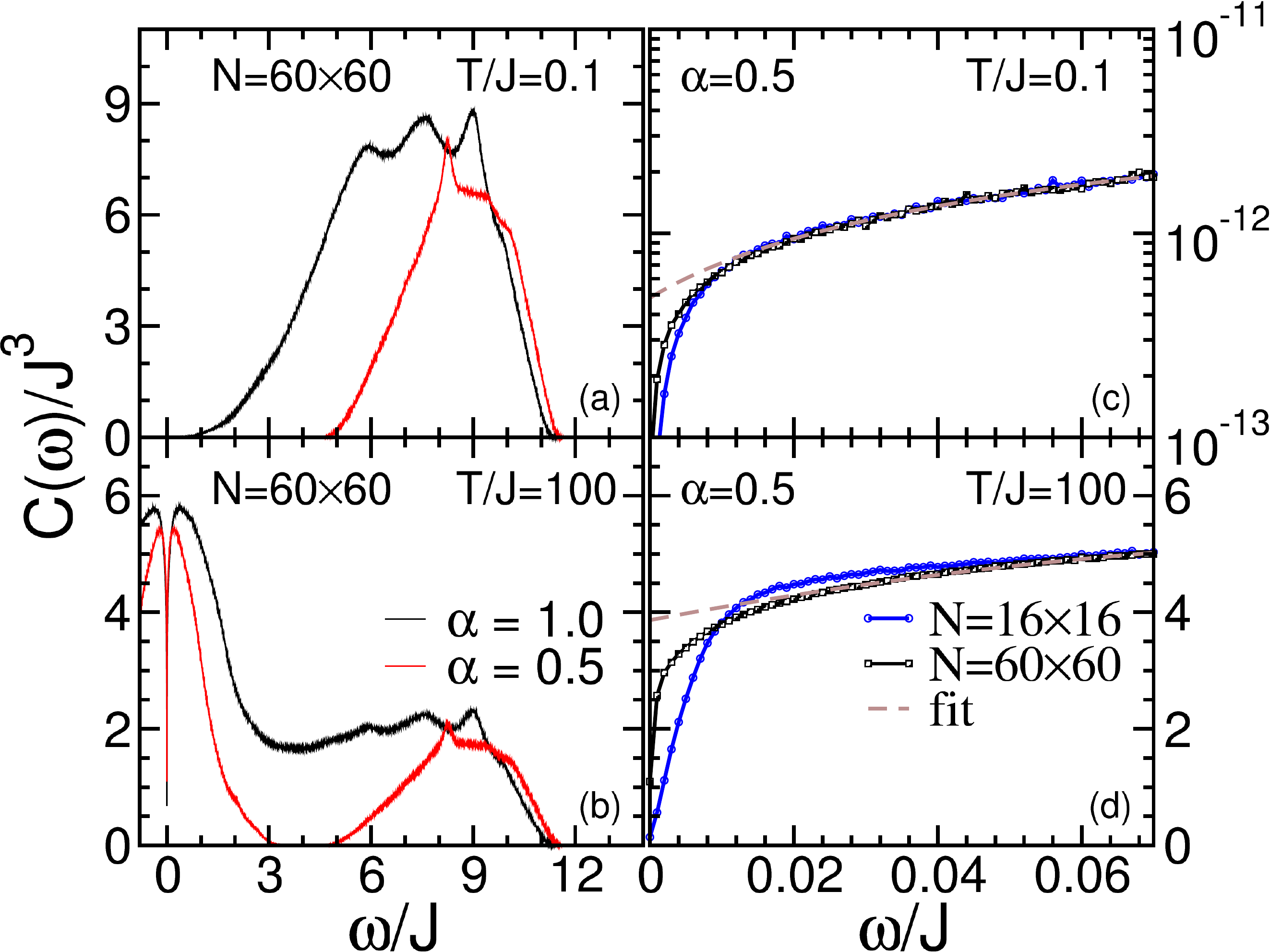}
\caption{ $C(\omega)$ from the AGC method, versus $\omega$ for two temperatures,
$T/J=0.1$ (a,c) and $100$ (b,d).  (a,b): Variation with anisotropy for
$\alpha=1, 0.5$ (black, red), corresponding to the gapless and gapful phase on a
large $\omega$-scale for $N{=}60{\times}60$ sites.  (c,d): Variation with system
size $N{=}16{\times}16, 60{\times}60$ (blue, black) at low-$\omega$ for
$\alpha=0.5$.  Brown dashed line: fitting-polynomial for $C(\omega,N{=}60 {\times} 60)$
to extract $\kappa^{dc}(T)$.}
\label{fig:CXX}
\end{figure}

Now we turn to the dynamical current correlation function $C(\omega)$. Results
on up to 60$\times$60-sites systems are displayed in Fig.~\ref{fig:CXX} for two
representative anisotropy values $\alpha=1.0(0.5)$. These place the system into
the gapless(ful) phase. Two temperatures, i.e. $T/J {=} 0.1, 100$ are
shown, which allow to highlight the difference between low and high
matter fermion densities. We begin with Fig. \ref{fig:CXX}(a).  Here the fermion
occupation number at all momenta is small and the intensity is essentially due
to two-fermion pair-breaking contributions to $C(\omega)$. For
$\alpha=0.5$, the one-particle dispersion displays a gap $\Delta$, at $\Delta /
J \approx 2$ \cite{Motome2015b}, with a corresponding two-particle excitation gap $\xi$ in
$C(\omega)$ at $\xi / J\approx 4$. Moving to Fig.~\ref{fig:CXX}(b) the
matter-fermion density is large and direct transport by thermally populated
quasiparticle states contributes. In a clean system this transport would show up
as a Drude peak strictly at $\omega{=}0$. Due to the gauge disorder however, the
Drude peak is smeared over a substantial range of finite frequencies $\sim
O(J)$. Interestingly, the randomness provided for by the gauge field is not
sufficient to fill in the gap completely, as is obvious from panel (a)
and (b). Related observations have been made in Refs.  \cite{Motome2015b,
Motome2016}. The 'smeared Drude peak' in Fig. \ref{fig:CXX}(b) features a very
narrow zero-frequency dip, which requires careful finite size analysis. Examples
of such analysis are shown in Fig. \ref{fig:CXX}(c,d) for the case of
$\alpha=0.5$. The main point can be read off best from panel (d), where it is
amply clear, that as $N\rightarrow\infty$ the zero-frequency dip closes in onto
the y-axis. This observation renders the system a normal, dissipative heat
conductor in the thermodynamic limit and extends identical findings from
Ref. \cite{Metavitsiadis2017} to the case of $\alpha\neq 1$.
Note, that there is a difference of several orders of magnitude 
between $C(\omega)$ in Figs. \ref{fig:CXX}(c) and \ref{fig:CXX}(d).
This is related to the thermal activation behavior in the gapful case,
to be discussed in the next section.

\subsection{DC-limit of heat conductivity}
\label{sub-secDCkappa}

\begin{figure}[tb]
\centering{}\includegraphics[width=0.98\columnwidth]{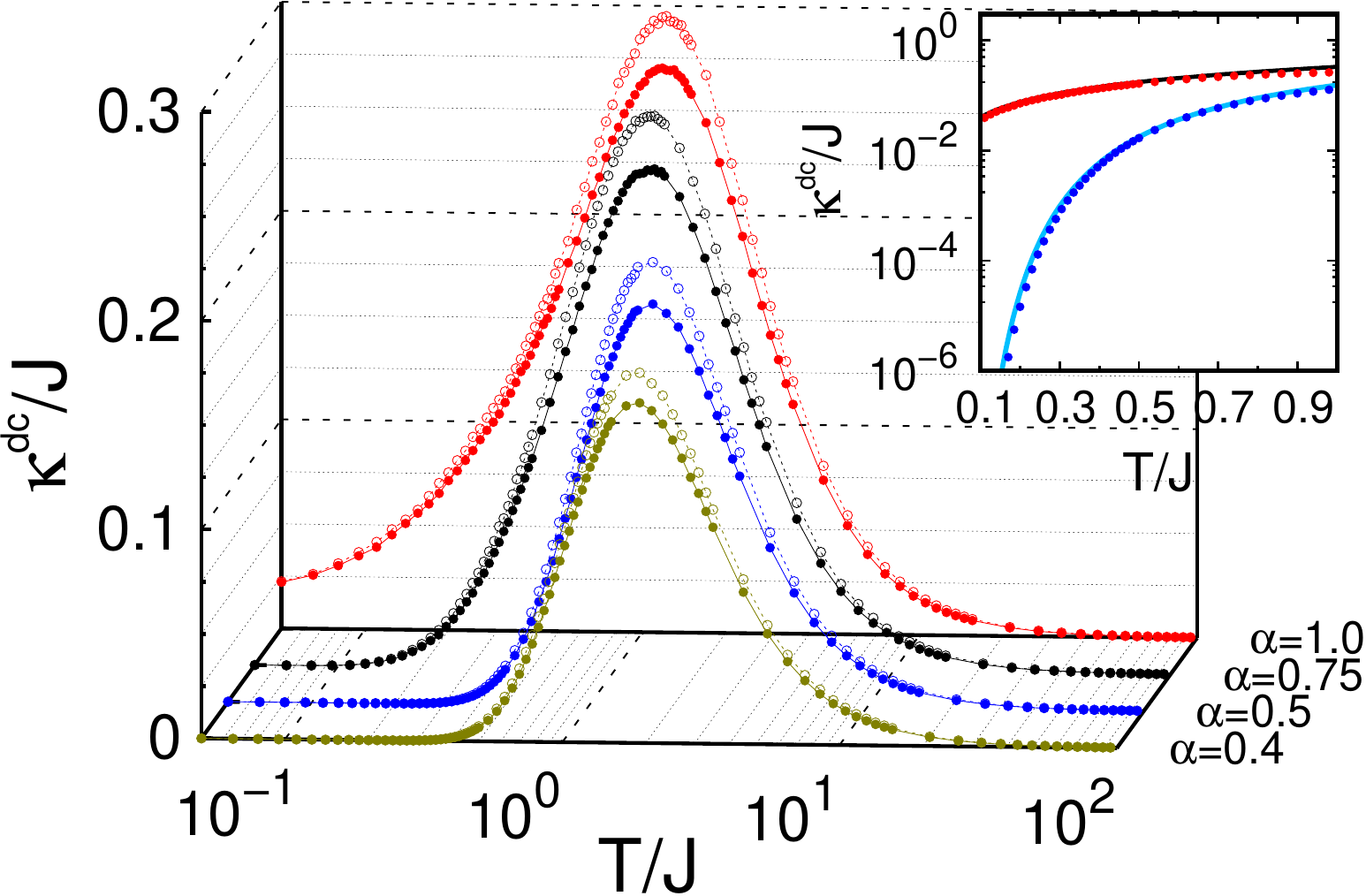}
\caption{
Thermal conductivity $\kappa^{dc}$ versus $T$ for various values of anisotropies
$\alpha=1.0 \dots 0.4$, shown on the lower right axis, ranging from the gapless
to gapped phases. Two system sizes, $N{=}16{\times}16$ (dashed with empty
circles), and $N{=}60{\times}60$ (solid with solid circles), are compared.
Inset: low-$T$ behavior for $\alpha=1.0$ (red dots) and $0.5$ (blue dots)
superimposed onto two distinct fit functions $T^\gamma$, with $\gamma{=}0.92$
(black solid) and $e^{-\Delta/T}$, with $\Delta{=}2.30$ (blue solid), for
$N{=}60{\times}60$.
}
\label{fig:kappa}
\end{figure}

Based on a series of calculations as in Fig.~\ref{fig:CXX}, we have extrapolated
the dc-limit $C(\omega \rightarrow 0,T,\alpha)$ over a wide range of $T$ and
$\alpha$, sufficient to acquire a consistent picture of the thermal conductivity
$\kappa^{dc}(T,\alpha)$ both, in the gapless and the gapful regime. This result
is depicted in Fig.~\ref{fig:kappa}. To arrive at this, the brown dashed lines
in Fig.~\ref{fig:CXX}(c,d) exemplarily indicate the extrapolation procedure
performed, for each of the largest systems, i.e. for $N{=}60{\times}60$.  It
consists of least-square fitting of the low-$\omega$ range of the correlation
function for each pair of $T,\alpha$ by second order polynomials, from which the
dc value $C(\omega \rightarrow 0, T)$ is extracted. We have chosen a range of
$0.02\leq\omega/J\leq 0.12$ to determine the coefficients of the polynomial in
all cases. To obtain some form of finite-size scaling, we have compiled the
results of the fitting procedure for two system sizes, i.e. $N{=}L{\times}L$ and
$L{=}\lbrace 16,60\rbrace$, into Fig.~\ref{fig:kappa}, showing dashed (solid)
curves with empty (solid) circles, for $L{=}16(60)$.  Although the linear
dimension of the systems shown differs by almost a factor of 4, $\kappa^{dc}$
displays only minimal differences for most of the temperatures. Only around
$T/J\approx 1$, a small, systematic correction towards lower values with
increasing system size is observed for every $\alpha$.  This implies visible,
but satisfyingly little finite-size effects, presumably putting the largest
system sizes rather close to the thermodynamic limit.

Fig.~\ref{fig:kappa} is a main result of this paper. It extends previous
analysis of the thermal conductivity of the Kitaev model \cite{Motome2017,
Metavitsiadis2017} into the anisotropic regime. It demonstrates a qualitative
change of the temperature dependence of $\kappa^{dc}(T,\alpha)$, as the system
crosses over from the gapless to the gapful spectrum. In the former the thermal
conductivity roughly scales with a power law, $\kappa^{dc}\sim T^\gamma$ for
$T/J\ll 1$, while in the latter an exponential behavior with $\kappa^{dc}\sim
\exp(-\Delta/T)$ provides a reasonable description. This is in line with the
notion of quasiparticle transport, since the activation gap $\Delta$ is
consistent with the one-particle gap, e.g. $\Delta/J\sim 2 $ for $\alpha=0.5$
as discussed in Sec. \ref{sub-secHeatCorr}.

We would like to emphasize, that while the trend of the curves in
Fig.~\ref{fig:kappa} might suggest a strictly vanishing $\kappa^{dc}$, e.g. for
$\alpha\geq 0.5$ and $T/J\lesssim 0.5$, this is {\it not} so, based on our
results. Rather, the y-axis scale in Fig.~\ref{fig:CXX}(c) is very small, albeit
finite. Such small orders of magnitude are consistent with the rapid low
temperature suppression by the exponential activation. In turn, our findings
suggest, that the Kitaev model is a normal dissipative heat conductor at all
finite temperatures, at least above $T^\star_\alpha$, for all $\alpha$.

\section{Conclusion}\label{secConcl}

In summary, we have investigated the longitudinal dynamical thermal conductivity
of the Kiteav model on the honeycomb lattice with anisotropic exchange
couplings. Using an average gauge configuration method, combined with exact
diagonalization of the matter fermion sector, systems of up 7200 spinful sites
have been analyzed over a wide range of temperatures and anisotropies. We have
confirmed both, the applicability and range of validity of our method by
complementary exact analysis of thermodynamic properties. Our main finding is
that for all temperatures and anisotropies investigated, the systems feature
normal dissipative transport, consistent with scattering of the matter fermions
from the static fluxes, and also consistent with a temperature dependence set by
the matter fermion mass gap. From an experimental point of view, transport in
putative Kitaev quantum spin liquids could be interesting under strain in order
to tune exchange couplings, while monitoring the temperature dependence of
thermal transport.

\section{Acknowledgments}

We thank M. Vojta and C. Hess for comments and discussion. Work of A.P. and
W.B. has been supported in part by the DFG through SFB 1143, project A02. Work
of W.B. has been supported in part by QUANOMET and the NTH-School CiN.
W.B. also acknowledges kind hospitality of the PSM, Dresden. This research was
supported in part by the National Science Foundation under Grant No. NSF
PHY-1748958.

\appendix
\numberwithin{equation}{section}

\begin{figure}[tb]
\centering{}\includegraphics[width=0.95\columnwidth]{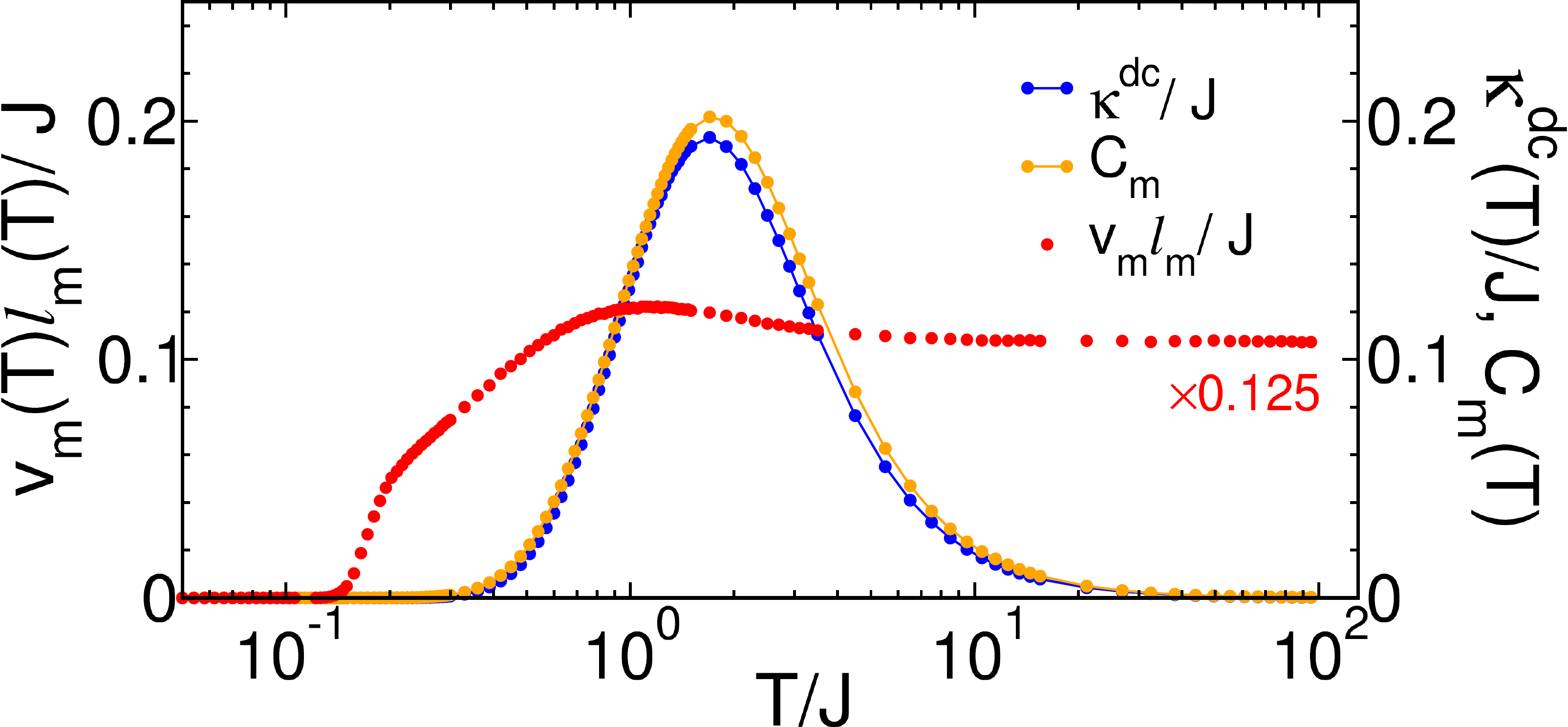}
\caption{Red solid circles: $\kappa^{dc}(T)/C_m(T)\equiv v_m(T) \ell_m(T)$.
versus $T$ at $\alpha=0.5$. Solid lines with solid circles: $\kappa^{dc} (T)$ (blue)
and $C_m(T)$ (yellow) evaluated by AGC method on $N=32\times32$ sites.}
\label{fig:MFPKT}
\end{figure}

\section{Kinetic model of thermal conductivity}\label{appKM}

This appendix is a digression into phenomenology, which we add out of curiosity.
It is customary to perform kinetic modeling of thermal
conductivities, expressing $\kappa^{dc}(T)$ in terms of quasiparticle properties
\cite{Hess2018a,Hess2001}
\begin{equation}\label{phenom}
\kappa^{dc}(T)\approx \sum_{k,p}C_{k,p}(T)v_{k,p}(T)\ell_{k,p}(T)\,,
\end{equation}
where $C_{k,p}(T), v_{k,p}(T)$, and $\ell_{k,p}(T)$ refer to the specific heat,
velocity, and mean free path at momentum $k$ of quasiparticles of type '$p$'. In
the present case these types are fluxes, $p{=}f$, and matter fermions,
$p{=}m$. Disentangling the contributions to $\kappa^{dc}(T)$ from various types
'$p$' is only an option, if their spectral supports are well separated. As one
can read off directly from Fig.~(\ref{fig:SHalpha1}), exactly this is possible
for strong anisotropy, e.g. $\alpha=0.5$, where the flux peak in $C_\nu(T)$ at
$T^\star_\alpha$ is well separated from the dominant contributions to the
specific heat of the fermions. Moreover, realizing that the flux velocity
$v_{k,f}=0$, and adopting the usual approximation to drop the momentum summation
in (\ref{phenom}) in favor of momentum integrated quantities, one gets
$\kappa^{dc}(T) \sim C_m(T) v_m(T) \ell_m(T)$ for $T>T^\star_\alpha$.

From this it is tempting to analyze the ratio of $\kappa^{dc}(T)/C_m(T)$, as
obtained from our calculations, e.g. for $\alpha=0.5$. This is shown in
Fig.~\ref{fig:MFPKT}. It is this variation of $v_m(T) \ell_m(T)$ with $T$, which
this appendix is meant to speculate on as follows. First, at all $T >
T^\star_\alpha$, the flux lattice is completely random, setting a mean free path
for the fermions, independent of $T$. Second, for sufficiently large $T$,
fermions far up in the Dirac cone set the Fermi velocity, which is also
independent of $T$. Third, upon lowering $T$, fermions close to the gap set
$v_m(T)$. This velocity approaches zero as $T\rightarrow 0$. Remarkably, these
points are roughly captured in the temperature variation of $v_m(T) \ell_m(T)$,
seen in Fig.~\ref{fig:MFPKT}.


\begin{thebibliography}{99}
\bibitem{noteSpec}
The specific heat is evaluated numerically via temperature derivative of entropy
\begin{equation}
S(T) = \frac{1}{T}(\langle E_{m}(T)\rangle - F(T))
\end{equation}
$E_{m}=\sum_{i}(\varepsilon_{i}^{\eta}-\frac{1}{2})a_i^\dag a_i$  
is the thermal energy of the matter fermions in a particular $\{\eta\}$
configuration, with $\varepsilon_{i}$ the single-particle energy, and 
$F(T) {=}-T\ln Z$ the free energy of the matter fermions. 
Here $Z=Tr_{\{\eta\}}Tr_{\{a(\eta)\}}e^{-\beta {\cal H}}$ is the total partition function.
\end{thebibliography}
\end{document}